\documentclass[twocolumn,showpacs,superscriptaddress,preprintnumbers,amsmath,amssymb]{revtex4}

\usepackage{graphicx}
\usepackage{dcolumn}
\usepackage{bm}
\usepackage{upgreek}

\begin{document}


\title{Electrothermal feedback in superconducting nanowire single-photon detectors}

\author{Andrew J. Kerman}
\affiliation{Lincoln Laboratory, Massachusetts Institute of
Technology, Lexington, MA, 02420}
\author{Joel K.W. Yang}
\affiliation{Research Laboratory of Electronics, Massachusetts
Institute of Technology, Cambridge, MA, 02139}
\author{Richard J. Molnar}
\affiliation{Lincoln Laboratory, Massachusetts Institute of
Technology, Lexington, MA, 02420}
\author{Eric A. Dauler}
\affiliation{Lincoln Laboratory, Massachusetts Institute of
Technology, Lexington, MA, 02420} \affiliation{Research Laboratory
of Electronics, Massachusetts Institute of Technology, Cambridge,
MA, 02139}
\author{Karl K. Berggren}
\affiliation{Research Laboratory of Electronics, Massachusetts
Institute of Technology, Cambridge, MA, 02139}

\date{\today}

\begin{abstract}
We investigate the role of electrothermal feedback in the operation
of superconducting nanowire single-photon detectors (SNSPDs). It is
found that the desired mode of operation for SNSPDs is only achieved
if this feedback is unstable, which happens naturally through the
slow electrical response associated with their relatively large
kinetic inductance. If this response is sped up in an effort to
increase the device count rate, the electrothermal feedback becomes
stable and results in an effect known as latching, where the device
is locked in a resistive state and can no longer detect photons. We
present a set of experiments which elucidate this effect, and a
simple model which quantitatively explains the results.
\end{abstract}

\maketitle

Superconducting nanowire single-photon detectors (SNSPDs)
\cite{summary,inductance,cavity,eric} have a combination of
attributes found in no other photon counter, including high speed,
high detection efficiency over a wide range of wavelengths, and low
dark counts. Of particular importance is their high
\textit{single-photon} timing resolution of $\sim$30 ps \cite{eric},
which permits extremely high data rates in communications
applications \cite{comm,QKD}. Full use of this electrical bandwidth
is limited, however, by the fact that the maximum count rates of
these devices are much smaller (a few hundred MHz for 10 $\mu$m$^2$
active area, and decreasing as the area is increased
\cite{inductance}), limited by their large kinetic inductance and
the input impedance of the readout circuit
\cite{inductance,parallel}. To increase the count rate, therefore,
one must either reduce the kinetic inductance (by using a smaller
active area or through the use of different materials or substrates)
or increase the load impedance \cite{parallel}. However, either one
of these approaches causes the wire to ``latch" into a stable
resistive state where it no longer detects photons \cite{joeltherm}.
This effect arises when negative electrothermal feedback, which in
normal operation allows the device to reset itself, is made fast
enough that it becomes stable. We present experiments which probe
the stability of this feedback, and we develop a model which
quantitatively explains our observations.

\begin{figure}
\includegraphics[width=3.25in]{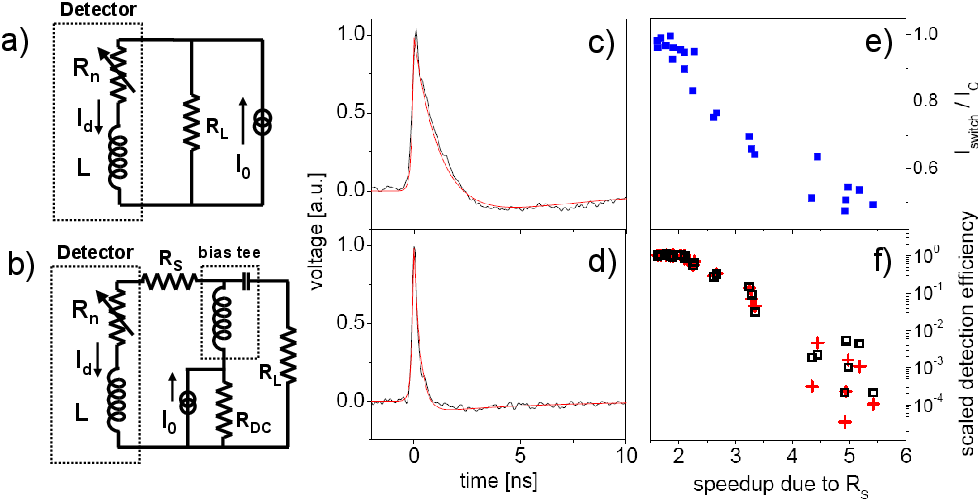}
\caption{Figure \ref{fig:1}:(color online) Speedup and latching of
nanowire detectors with increased load impedance. (a) electrical
model of detector operation. A hotspot is nucleated by absorption of
a photon, producing a resistance $R_n$ in series with the wire's
kinetic inductance $L$. (b) circuit describing experimental
configuration for increasing load resistance using a series resistor
$R_S$. (c) and (d) averaged pulse shapes from $3\mu$m$\times
3.3\mu$m detectors with $L\sim50$ nH for $R_S=0$ and
$R_S=250\Omega$; red solid lines are predictions with no free
parameters. (e) switching current
$I_{switch}\equiv\textrm{min}(I_c,I_{latch})$ vs. speedup
$50\Omega/(R_S+50\Omega)$. For small $R_S$, $I_\textrm{switch}=I_c$
but as $R_S$ is increased, $I_\textrm{latch}$ decreases, becoming
less than $I_c$. (f) DE at $I_0=0.975I_\textrm{switch}$ vs. speedup
(open squares). Also shown (crosses) are the expected DEs assuming
that latching affects DE simply by limiting $I_0$ (obtained from DE
vs. $I_0$ at $R_S=0$).~\label{fig:1}}
\end{figure}

The operation of an SNSPD is illustrated in Fig.~\ref{fig:1}(a). A
nanowire (typically $\sim$100nm wide, 5nm thick) is biased with a DC
current $I_0$ near the critical current $I_c$. When a photon is
absorbed, a short ($<$100 nm long) normal domain is nucleated,
giving the wire a resistance $R_n(t)$; this results in Joule heating
which causes the normal domain (and consequently, $R_n$) to expand
in time exponentially. The expansion is counteracted by negative
electrothermal feedback associated with the load $R_L$ (typically a
$50\Omega$ transmission line), which forms a current divider with
the nanowire and tends to divert current out of it, thereby reducing
the heating (and producing a voltage across the load). In a
correctly functioning device, this feedback is sufficiently slow so
that it is unstable: that is, the inductive time constant is long
enough so that before appreciable current is diverted out of the
load, Joule heating has already caused the normal domain to grow
large enough so that $R_n\gg R_L$ \cite{size}. The current in the
device then drops nearly to zero, allowing it to quickly cool down
and return to the superconducting state, after which the current
recovers with a time constant $\tau_e\equiv L/R_L$
\cite{inductance}. If one attempts to shorten $\tau_e$, at some
point the negative feedback becomes fast enough to counterbalance
the fast Joule heating before it runs away, resulting in a stable
resistive domain, known as a self-heating hotspot \cite{GM}.

In a standard treatment of these hotspots \cite{GM}, solutions to a
one-dimensional heat equation are found in which a
normal-superconducting (NS) boundary propagates at constant velocity
$v_{ns}$, for fixed device current $I_d$ \cite{GM,TC}. This results
in a solution of the form:

\begin{eqnarray}
v_{ns}=v_0\frac{\alpha(I_d/I_c)^2-2}{\sqrt{\alpha(I_d/I_c)^2-1}}\approx\frac{1}{\gamma}(I_d^2-I_{ss}^2)\label{eq:vss}
\end{eqnarray}

\noindent where $v_0\equiv\sqrt{A_{cs}\kappa h}/c$ is a
characteristic velocity ($A_{cs}$ is the wire's cross-sectional
area, $\kappa$ is its thermal conductivity, and $c$ and $h$ are the
heat capacity and heat transfer coefficient to the substrate, per
unit length, respectively), $I_c$ is the critical current, and
$\alpha\equiv\rho_nI_c^2/h(T_c-T_0)$ is known as the Stekly
parameter ($\rho_n$ is the normal resistance of the wire per unit
length, $T_0$ and $T_c$ are the substrate and superconducting
critical temperatures) which characterizes the ratio of joule
heating to conduction cooling in the normal state \cite{GM}.
Equation \ref{eq:vss} is valid when $T_0\ll T_c$, and the
approximate equality holds for small deviations $\delta I\equiv
I-I_{ss}$, with $\gamma\equiv(T_c-T_0)(c/\rho_n)\sqrt{h/\kappa
A_{cs}}$ and $I_{ss}^2\equiv 2h(T_c-T_0)/\rho_n$. The physical
meaning of eq. \ref{eq:vss} is clear: the NS boundary is stationary
only if the local power density is equal to a fixed value; if it is
greater, the hotspot will expand ($v_{ns}>0$), if less it will
contract ($v_{ns}<0$).

We can use eq. \ref{eq:vss} to describe the electrothermal circuit
in fig.~\ref{fig:1}(a), by combining it with:
$dR_n/dt=2\rho_nv_{ns}$ and $I_dR_n+LdI_d/dt=R_L(I_0-I_d)$. Assuming
that the small-signal stability of the DC hotspot solution
($I_d=I_{ss}, R_{ss}=R_L(I_0/I_{ss}-1)$) determines whether it will
latch, we linearize about this solution to obtain a second-order
system, with damping coefficient
$\zeta=\frac{I_0}{4I_{ss}}\sqrt{\tau_{th}/\tau_{e}}$, where
$\tau_{th}\equiv R_L/2\rho_nv_0$ is a thermal time constant. This
can be re-expressed as
$\zeta=\frac{1}{4}\sqrt{\tau_{th,tot}/\tau_{e,tot}}$, with
$\tau_{e,tot}\equiv L/R_{tot}$ and $\tau_{th,tot}\equiv
R_{tot}/2\rho_nv_0$, where $R_{tot}\equiv R_L+R_{ss}$. If the
damping is less than some critical, minimum value $\zeta_{latch}$,
the feedback cannot stabilize the hostpot during the initial
photoresponse, as described above, and the device operates normally.
However, since the steady-state solution gives $R_{ss}\propto I_0$
(as $I_0$ is increased, a larger hotspot is necessary for
$I_d=I_{ss}$) the hotspot becomes more stable as $I_0$ is increased,
until eventually $\zeta(I_{latch})=\zeta_{latch}$ and the device
latches. For a correctly functioning device, $I_{latch}>I_c$, so
that latching does not affect its operation. However, if $\tau_e$ is
decreased, $I_\textrm{latch}$ decreases, and eventually it becomes
less than $I_c$. This prevents the device from being biased near
$I_c$ \cite{biastee}, resulting in a drastic reduction in
performance \cite{constriction, jitter}.

Devices used in this work were fabricated from $\sim$5 nm thick NbN
films, deposited on R-plane sapphire substrates in a UHV DC
magnetron sputtering system (base pressure $<10^{-10}$ mbar). Film
deposition was performed at a wafer temperature of $\sim$800 C and a
pressure of $\sim 10^{-8}$ mbar \cite{molnar}. Aligned
photolithography and liftoff were used to pattern $\sim$100 nm thick
Ti films for on-chip resistors \cite{joeltherm}, and Ti:Au contact
pads. Patterning of the NbN was then performed with e-beam
lithography using HSQ resist \cite{cavity}. Devices were tested in a
cryogenic probing station at temperatures of 1.8-12K using the
techniques described in Refs. \cite{inductance,cavity}.

Figure \ref{fig:1}(c)-(f) show data for a set of (3$\mu$m$\times
3.3\mu$m active area) devices having various resistors $R_S$ in
series with the 50$\Omega$ readout line \cite{joeltherm}
[fig.~\ref{fig:1}(b)] so that $R_L=50\Omega+R_S$. Panels (c) and (d)
show averaged pulse shapes for devices with $R_S=0$ and
$R_S=250\Omega$, respectively. Clearly, the reset time can be
reduced; however, this comes at a price. Panels (e) and (f) show,
for devices with different $R_S$, the current
$I_\textrm{switch}\equiv\textrm{min}(I_c,I_{latch})$ above which
each device no longer detects photons, and the measured detection
efficiency (DE) at $I_0=0.975I_{switch}$ \cite{biastee}. The data
are plotted vs. the speedup of the reset time
$(R_S+50\Omega)/50\Omega$, and show that as this speedup is
increased, $I_{switch}$ decreases (due to reduction of $I_{latch}$),
resulting in a significantly reduced DE.

\begin{figure}
\includegraphics[width=3.25in]{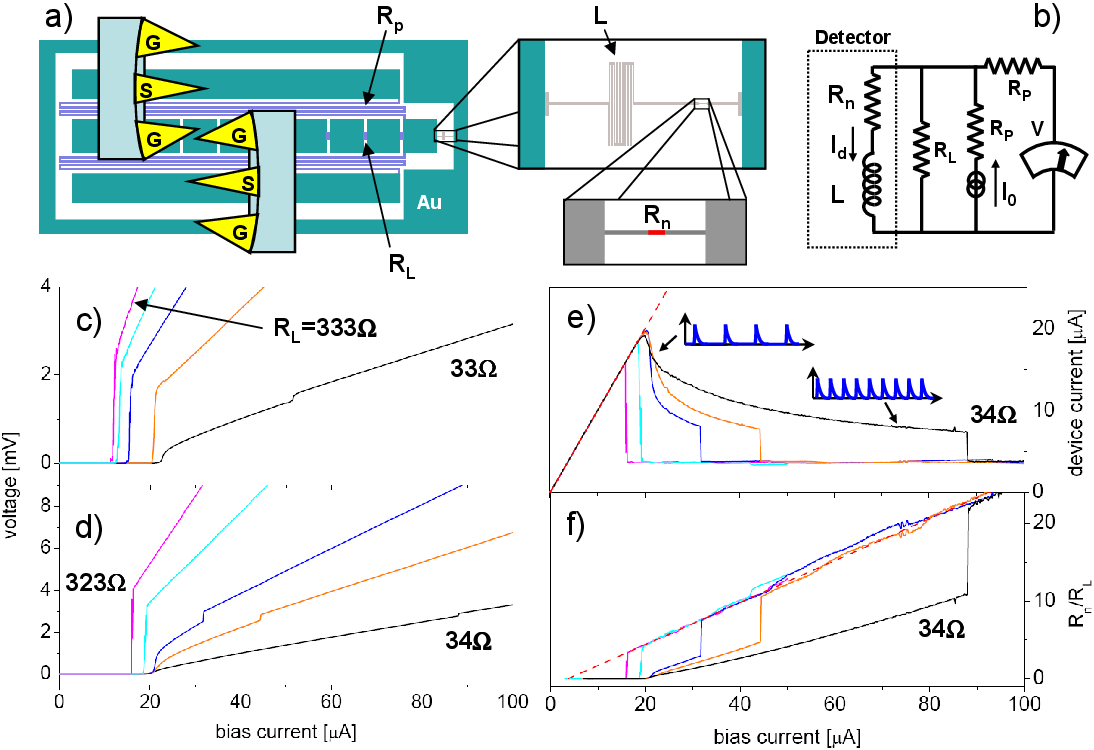}
\caption{Figure \ref{fig:2}: (color online) Hotspot stability
measurements. (a) device schematic; two ground-signal-ground probes
simultaneously perform a high-impedance 3-point measurement of the
hotspot resistance, with $R_L$ determined by probe position. (b)
equivalent electrical circuit. (c) and (d) example V-I curves, with
$L=60$ nH and $L=605$ nH, respectively; (e) and (f) inferred $I_d$
and $R_n/R_L$ for the data shown in (d). In (e) the relaxation
oscillations in the region $I_c<I_0<I_\textrm{latch}$ are shown
schematically. Dashed lines show (e) $I_d=I_0$ and (f)
$R_{ss}=R_L(I_0/I_{ss}-1).$\label{fig:2}}
\end{figure}

To investigate the latched state, we fabricated devices designed to
probe the stability of self-heating hotspots as a function of $I_0$,
$L$, and $R_L$. Each device consisted of three sections in series,
as shown in Fig.~\ref{fig:2} (a): a 3 $\mu$m-long, 100 nm-wide
nanowire where the hotspot was nucleated \cite{size}; a wider (200
nm) meandered section acting as an inductance; and a series of nine
contact pads interspersed with Ti-film resistors. Also shown are the
two electrical probes, which result in the circuit of
Fig.~\ref{fig:2}(b): a high-impedance ($R_p=20\textrm{k}\Omega$)
3-point measurement of the nanowire resistance. We varied $R_L$ by
touching the probes down to different pads along the line, and L by
testing different devices (with different series inductors). We
tested 66 devices on three chips, and selected from these only
unconstricted \cite{constriction} wires with nearly identical
linewidths (the observed $I_c$ of devices used here were within
$\sim$10\% of each other - typically 22-24 $\mu$A), spanning the
range: $R_L=20-1000\Omega$ and $L=6-600$ nH.

For each $L$ and $R_L$, we acquired a DC I-V curve like those shown
in Figs.~\ref{fig:2}(c),(d), sweeping $I_0$ downward starting from
high values where the hotspot was stable \cite{upward}. All curves
exhibit a step in voltage which can be identified as
$I_\textrm{latch}$. For smaller $R_L$, where $I_\textrm{latch}$ is
large, $I_c$ can be identified as the point where an onset of
resistance occurs common to several curves; this onset is more
gradual with $I_0$ rather than sudden as at $I_\textrm{latch}$.
These features can be understood by examining figs.~\ref{fig:2}(e)
and (f), which show $I_d$ and $R_n$ inferred from the data of
fig.~\ref{fig:2}(d). Above the discontinuous jump in current at
$I_\textrm{latch}$, $I_d$ is fixed at $I_{ss}$ (independent of $I_0$
and $R_L$) indicating the latched state. For smaller $\tau_e$,
$I_\textrm{latch}<I_c$, so only $I_\textrm{latch}$ is observed
\cite{retrap}. When $\tau_e$ is large enough that
$I_\textrm{latch}>I_c$, an intermediate region appears where the
resistance increases continuously with $I_0$; this arises from
relaxation oscillations \cite{hadfield,GM}, as indicated in the
figure: the device cannot superconduct when $I_0>I_c$, but neither
can a stable hotspot be formed when $I_0<I_\textrm{latch}$, so
instead current oscillates back and forth between the device and the
load, producing a periodic pulse train with a frequency that
increases as $I_0$ is increased \cite{biastee}. The average
resistance increases with this frequency, producing the observed
continuous decrease in $I_d$.

\begin{figure}
\includegraphics[width=3.25in]{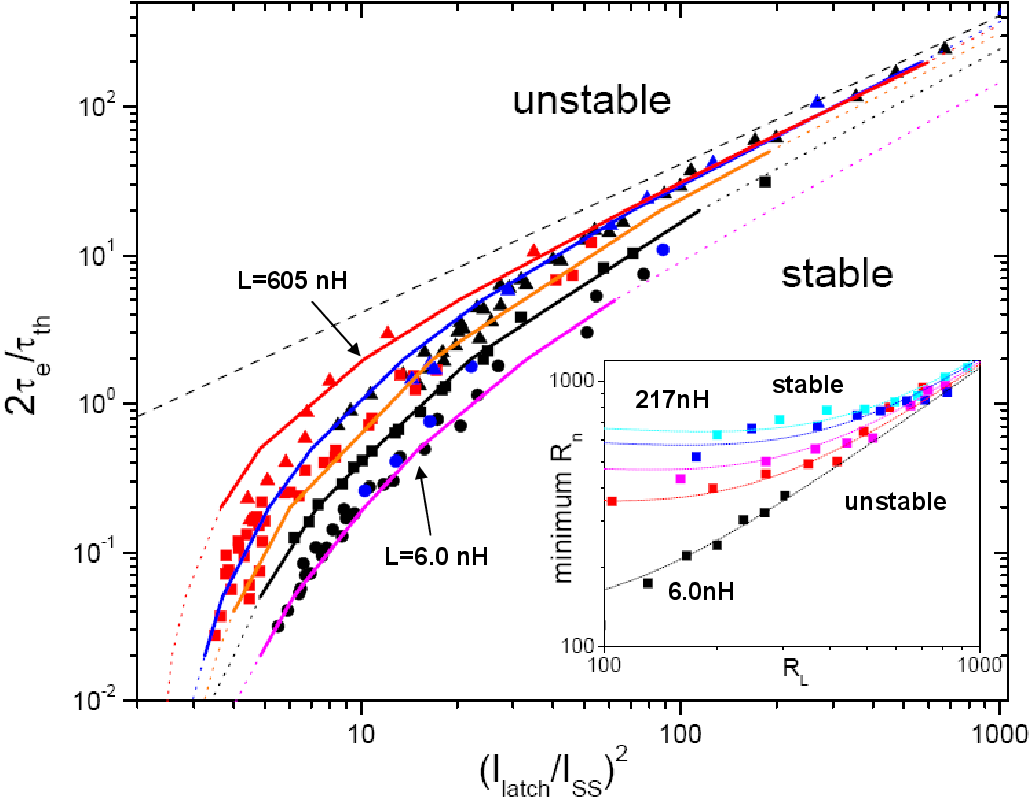}
\caption{Figure \ref{fig:3}: (color online) Summary of hotspot
stability results. Data are shown from 3 different chips (indicated
by different colors). Circles, squares, and triangles are data for
$L=$6-12, 15-60, and 120-600 nH, respectively. In the
$\tau_e\gg\tau_a$ limit (where NS domain-wall motion dominates the
electrothermal feedback), the data approach the dashed line, which
is the prediction based on eq. \ref{eq:vss}, discussed in the text.
For $\tau_e\ll\tau_{th}$ the NS domain walls are effectively fixed
and the temperature feedback dominates. In this regime the feedback
is always unstable when $R_{ss}\sim >R_L$ (or equivalently, $I_0\sim
<2I_{ss}$), as shown in the inset. The solid curves are obtained
from eq.~\ref{eq:aol} assuming a phase margin of 30 degrees; each
curve corresponds to a fixed $L$ in the set (6,15,30,60,600) nH and
spans the range of $R_L$ in the data. The dotted lines extend these
predictions over a wider range of $R_L$.\label{fig:3}}
\end{figure}

Figure \ref{fig:3} shows the measured $I_{latch}$ as a function of
$R_L$ and $L$, which can be thought of as defining the boundary
between stable and unstable hotspots. Our simple model described
above predicts:
$\tau_e/\tau_{th}\propto(I_\textrm{latch}/I_{ss})^2$, a line of
slope 1 in the figure (indicated by the dashed line). The data do
approach this line, though only in the $\tau_e\gg\tau_{th}$ limit.
This is consistent with the assumption of constant (or slowly
varying) $I_d$ under which eq.~\ref{eq:vss} was derived. As
$\tau_e/\tau_{th}$ is decreased, the data trend downward, away from
this line, and $I_\textrm{latch}/I_{ss}$ becomes \textit{almost
independent of} $\tau_e/\tau_{th}$; this implies a minimum
$I_\textrm{latch}/I_{ss}$, or equivalently, a minimum $R_{ss}/R_L$,
below which the hotpost is \textit{always} unstable. This is shown
in the inset: the measured minimum stable $R_{ss}$ is always greater
than $\sim R_L$.

The crossover to this behavior can be explained in terms of a
timescale $\tau_a$ over which the temperature profile of the hotspot
stabilizes into the steady-state form which yields eq.~\ref{eq:vss}.
For power density variations occuring faster than this, the NS
boundaries do not have time to start moving, resulting instead in a
temperature deviation $\Delta T$. Since the NS boundary occurs at
$T\approx T_c$, where $\rho_n$ is temperature-dependent (defined by
$d\rho/dT\equiv\beta >0$), this changes $R_n$, giving a second,
parallel electrothermal feedback path which dominates for
frequencies $\omega\gg\tau_a^{-1}$. We can describe this by
replacing eq.~\ref{eq:vss} with:

\begin{equation}
\frac{\gamma\rho_n}{2}\left
(\tau_a\frac{d^2l}{dt^2}+\frac{dl}{dt}\right )=I_d^2\rho(\Delta
T)-I_{ss}^2\rho_n\label{eq:l}
\end{equation}

\begin{equation}
c\frac{d\Delta
T}{dt}=\frac{\gamma\rho_n\tau_a}{2}\frac{d^2l}{dt^2}-h\Delta
T\label{eq:t}
\end{equation}

\noindent Here, $l$ is the hotspot length, $\rho(\Delta T)$ is the
resistance per unit length, and $R_n=\rho(\Delta T)l$. In eq.
\ref{eq:l}, $\tau_a$ is the characteristic time over which
$dl/dt=2v_{ns}$ adapts to changes in power density: for slow
timescales $dt\gg\tau_a$, $\tau_ad^2l/dt^2\ll dl/dt$ and eq.
\ref{eq:l} reduces to eq. \ref{eq:vss} (with $\rho=\rho_n$). For
faster timescales, $\tau_ad^2l/dt^2$ becomes appreciable, and acts
as a source term for temperature deviations in eq. \ref{eq:t}. When
$dt\ll\tau_a$, $\tau_ad^2l/dt^2\gg dl/dt$ and eqs.~\ref{eq:l}
and~\ref{eq:t} can be combined to give: $cd\Delta T/dt\approx
I_d^2\rho-I_{ss}^2\rho_n-h\Delta T$ \cite{tes}. In this limit, if
$R_L\sim R_{ss}$ the bias circuit including $R_L$ begins to look
like a current source, which then results in positive feedback: a
current change produces a temperature and resistance change of the
same sign. Therefore, the hotspot is always unstable when
$R_{ss}<~R_L$.

Expressing eqs.~\ref{eq:l} and \ref{eq:t} in dimensionless units
($i\equiv I_d/I_0$, $r\equiv R_n/R_L$, $\lambda\equiv l\beta
T_c/R_L$, $\theta\equiv T/T_c$), and expanding to first order in
small deviations from steady state, we obtain:




\begin{equation}
\delta i^\prime=-\left (i_0\delta i+i_0^{-1}\delta r\right
)\label{eq:1}
\end{equation}
\begin{equation}
\delta r=\eta \left(i_0-1\right)\delta\theta+\eta^{-1}\delta\lambda
\end{equation}
\begin{equation}
\frac{\tau_a}{\tau_e}\delta\lambda^{\prime\prime}+\delta\lambda^\prime=2\eta^2
\frac{\tau_e}{\tau_{th}}\left ( \delta\theta+2i_0\eta^{-1}\delta
i\right )
\end{equation}
\begin{equation}
\delta\theta^\prime=\frac{\Theta\tau_{th}\tau_a}{\eta
\tau_e\tau_c}\delta\lambda^{\prime\prime}-\frac{\tau_e}{\tau_c}\delta\theta\label{eq:4}
\end{equation}

\noindent Here, the prime denotes differentiation with respect to
$t/\tau_e$, $i_0\equiv I_0/I_{ss}$, $\Theta\equiv (T_c-T_0)/T_c$,
$\eta\equiv\beta T_c/\rho_n$ characterizes the resistive transition
slope, and $\tau_c\equiv c/h$ is a cooling time constant. When
$\tau_e\gg\tau_{th},\tau_a$, the system reduces to: $\delta
i^{\prime\prime}+i_0\delta i^\prime-4\tau_e/\tau_{th}\approx 0$,
which has damping coefficient
$\zeta=i_0(4\sqrt{\tau_e/\tau_{th}})^{-1}$, as above. In the
opposite limit, where $\tau_e\ll\tau_{th},\tau_a$, we obtain:
$\delta i^{\prime\prime}+i_0\delta
i^\prime+(2\eta\Theta\tau_e/\tau_c)(i_0-2)\approx 0$. In agreement
with our argument above, the oscillation frequency becomes negative
for $R_{ss}<R_L$ ($I_0<2I_{ss}$).

We characterize the stability of the system of eqs.
\ref{eq:1}-\ref{eq:4} using its ``open loop" gain $A_{ol}$: we
assume a small oscillatory perturbation by replacing $\delta r$ in
eq.~\ref{eq:1} with $\Delta re^{j\omega t}$, and responses $(\delta
i, \delta\theta, \delta\lambda,\delta r)e^{j\omega t}$. Solving for
$A_{ol}\equiv\delta r/\Delta r$, we obtain:

\begin{equation}
A_{ol}=\frac{4\frac{\tau_e}{\tau_{th}}(1+j\omega\frac{\tau_c}{\tau_e})-4\eta
\Theta\omega^2(i_0-1)\frac{\tau_a}{\tau_e}}{j\omega
i_0(1+\frac{j\omega}{i_0})
\left[2j\omega\eta\Theta\frac{\tau_a}{\tau_e}-(1+j\omega
\frac{\tau_c}{\tau_e})(1+j\omega\frac{\tau_a}{\tau_e})\right]}\label{eq:aol}
\end{equation}

\noindent The stability of the system can then be quantified by the
phase margin: $\pi$+arg[$A_{ol}(\omega_0)$], where $\omega_0$ is the
unity gain ($|A_{ol}|=1$) frequency. In the extreme case, when the
phase margin is zero (arg[$A_{ol}(\omega_0)]=-\pi$), the feedback is
positive. The solid lines in Fig. \ref{fig:3} show our best fit to
the data. Note that although the stability is determined only by
$\tau_e/\tau_{th}$ and $i_0$ in the two extreme limits (not visible
in the figure), in the intermediate region of interest here this is
not the case, so several curves are shown. Each solid curve segment
corresponds to a single $L$, over the range of $R_L$ tested; the
dotted lines continue these curves for a wider range of $R_L$. The
data are grouped into three inductance ranges: 6-12, 15-60, and
120-600 nH, indicated by circles, squares, and triangles,
respectively. We used fixed values $\Theta=0.8$, $\eta=6.5$, which
are based on independent measurements, and fitted $\tau_a=1.9$ ns,
and $\tau_c=7.7$ ns to all data. Separate values of $\rho_nv_0$ were
fitted to data from each of the three chips, differing at most by a
factor of $\sim$2. These fitted values were $\rho_nv_0\sim 1\times
10^{11}\Omega$/s; since $\rho_n\sim 10^9\;\Omega$/m, this gives
$v_0\sim 100$ m/s, a reasonable value.

A natural question to ask in light of this analysis is whether it
suggests a method for speeding up these devices. The most obvious
way would be to increase the heat transfer coefficient $h$, which
increases both $I_{ss}$ and $v_0$, moving the wire further into the
unstable region, and allowing its speed to be increased further
without latching. However, at present it is unknown how much $h$ can
be increased before the DE begins to suffer. At some point, the
photon-generated hotspot will disappear too quickly for the wire to
respond in the desired fashion. In any case, experiments like those
described here will be a useful measurement tool in future work for
understanding the impact of changes in the material and/or substrate
on the thermal coupling and electrothermal feedback.

We acknowledge helpful discussions with Sae Woo Nam, Aaron Miller,
Enectal\'i Figueroa-Feliciano, and Jeremy Sage.

This work is sponsored by the United States Air Force under Contract
\#FA8721-05-C-0002. Opinions, interpretations, recommendations and
conclusions are those of the authors and are not necessarily
endorsed by the United States Government.


\begin{thebibliography}{}
\bibitem{TC} This description is further simplified by the fact that near the
NS boundary all material properties can be approximated by their
values at $T_c$.
\bibitem{size} The sheet resistance of the NbN is $R_\square\approx
500\Omega$, and that the longest hotspot measured in this work is
$\sim$300 nm.
\bibitem{jitter} We have focused on the impact of electrothermal feedback
on reset time, but it may also influence the timing jitter since it
opposes the fast, initial growth of the normal domain that produces
the sharp leading-edge of the output pulses.
\bibitem{upward} The results are almost identical when sweeping
$I_0$ upward, since the dark counts of the device allow it to lock
into the latched state if it is stable.
\bibitem{biastee} In typical experiments, the DC impedance to ground
$R_{DC}$ is determined by the $I_0$ source behind a bias tee,
[Fig.~\ref{fig:1}(b) with $R_S=0$]. When $I_c$ is exceeded, the
detector oscillates (if $I_\textrm{latch}>I_c$), for the time
constant of the bias tee, after which it senses the larger $R_{DC}$
and latches. The absence of this burst of pulses is a signature for
$I_\textrm{latch}<I_c$.
\bibitem{retrap} When $I_\textrm{latch}<I_c$, $I_{latch}$ is
equivalent to what is often called the ``retrapping" current in
superconducting devices exhibiting self-heating.
\end{thebibliography}
\end{document}